\documentstyle[twocolumn,aps,prd]{revtex}
\tighten
\begin{document}
\draft

\title{Explicit Gravitational Radiation in Hyperbolic Systems
for Numerical Relativity}
\author{C.~Bona, C.~Palenzuela}
\address{Departament de Fisica. Universitat de les Illes Balears.
     Ctra de Valldemossa, km 7.5. 07071 Palma de Mallorca. Spain.}

\date{7 February 2002}

\maketitle

\begin{abstract}
A method for studying the causal structure of space-time evolution
systems is presented. This method, based on a generalization of
the well known Riemann problem, provides intrinsic results which
can be interpreted from the geometrical point of view. A
one-parameter family of hyperbolic evolution systems is presented
and the physical relevance of their characteristic speeds and
eigenfields is discussed. The two degrees of freedom corresponding
to gravitational radiation are identified in an intrinsic way,
independent of the space coordinate system. A covariant
interpretation of these degrees of freedom is provided in terms of
the geometry of the wave fronts. The requirement of a consistent
geometrical interpretation of the gravitational radiation degrees
of freedom is used to solve the ordering ambiguity that arises
when obtaining first order evolution systems from the second order
field equations. This achievement provides a benchmark which can
be used to check both the existing and future first order
hyperbolic formalisms for Numerical Relativity.
\end{abstract}

\section{Introduction}

The structure of Einstein field equations has deserved great
interest since the very beginning of General Relativity. It was
early noticed that, by rearranging the order or partial
derivatives, the principal part of the four dimensional Ricci
tensor could be written as a sort of generalized wave equation
\cite{DeDonder,Lanczos}
\begin{equation}\label{Ricci4D}
   2\;R_{\mu\nu}=-\Box g_{\mu\nu} + \partial_{\mu}\Gamma_{\nu}
                 + \partial_{\nu}\Gamma_{\mu}+...
\end{equation}
where the box stands for the d'Alembert operator acting on
functions and we have written for short
\begin{equation}\label{Gamma4D}
   \Gamma^{\mu}\equiv g^{\rho\sigma}\;{\Gamma^{\mu}}_{\rho\sigma}
    = -\Box x^{\mu}\;\;.
\end{equation}

This opened the way to the use of harmonic spacetime coordinates
($\Box x^{\mu}=0$) in order to obtain an hyperbolic evolution
system \cite{Choquet55,Hawking,DeTurck,Friedrich85}.

By the middle of the past century, however, the interest focused
in the relativistic Cauchy problem. The 3+1 decomposition of the
line element:
\begin{eqnarray}\label{metric4D}
    ds^2 &=& - \alpha^2\;dt^2  \nonumber \\
    &+& \gamma_{ij}\;(dx^i+\beta^i\;dt)\;(dx^j+\beta^j\;dt)
     \;\;\;\;i,j=1,2,3
\end{eqnarray}
allowed one to express six of the ten second order original
equations as a system of evolution equations for the metric
$\gamma_{ij}$ and the extrinsic curvature $K_{ij}$ of the
$t=constant$ slices, namely:
\begin{eqnarray}
   (\partial_t -{\cal L}_{\beta})\gamma_{ij} &=& -2\;\alpha\;K_{ij}
\label{evolve_metric} \\
   (\partial_t -{\cal L}_{\beta}) K_{ij} &=& -\nabla_i\alpha_j
       + \alpha\; [{}^{(3)}R_{ij}-2K^2_{ij}+ trK\;K_{ij}]
\label{evolve_curv}
\end{eqnarray}
where $\cal L $ stands for the Lie derivative (we restrict
ourselves to the vacuum case for simplicity). The remaining four
equations could instead be expressed as constraints:
\begin{eqnarray}
   ^{(3)}R - tr(K^2) + (trK)^2 &=& 0
\label{energy_constraint}  \\
   \nabla_k\:{K^k}_{i}-\partial_i(trK) &=& 0 \;\;\;.
\label{momentum_constraint}
\end{eqnarray}

This opened the door to a new way of obtaining hyperbolic
evolution systems \cite{Choquet83,Bona92,Frittelli94}. The key
point was to use in one or another way the momentum constraint
(\ref{momentum_constraint}) to ensure hyperbolicity while keeping
the freedom of choosing arbitrary space coordinates on every
$t=constant$ slice. This allows for instance to use normal
coordinates ($\beta^{i}=0$) without affecting the mathematical
structure of the evolution system.(See Ref. \cite{Bona99} for a
detailed comparison of the "old" and "new" hyperbolic systems).

These findings came at the right moment for people working on
Numerical Relativity with a view on the gravitational waves
detector projects starting by the turn of the century. Following
the wake of these first works, many groups found their own way of
combining the momentum constraint with the evolution equation
(\ref{evolve_curv}), leading in each case to a new brand of
hyperbolic systems
\cite{Choquet95,Bona95,Bona97,Frittelli96,Friedrich96,Putten96,Estabrook97,Iriondo97,Bonilla98,Stewart98,Yoneda99,Yoneda00,Yoneda01,Alcubierre99,Anderson99,Cornell01}.

Suddenly, the problem of having too few hyperbolic formalisms for
Numerical Relativity turned into the opposite problem of having
too many of them. Some of the works considered even multiparameter
families of hyperbolic systems
\cite{Frittelli96,Bona99,Cornell01}. Faced with the problem of
choice, we think that the right question at this point is: what
are hyperbolic systems for?

There are many answers, of course, but we can get a  hint from the
physical point of view when we realize that all but one of the
resulting characteristic speeds can be altered when playing with
these arbitrary parameters (see the work of the Cornell group
\cite{Cornell01} for a fairly complete analysis). The only one
that has an intrinsic value (light speed) is associated with the
two (transverse traceless) degrees of freedom of gravitational
waves. The evident question now is wether or not we can directly
relate the corresponding explicit eigenfields with gravitational
radiation propagation. We will see below that a positive answer
actually determines one of the parameters, introduced in the
Cornell paper \cite{Cornell01}. We claim that the requirement of
having explicit eigenvectors describing the gravitational waves
degrees of freedom in an intrinsic way (for an arbitrary choice of
space coordinates) is a strong benchmark to discriminate between
different hyperbolic evolution systems.

In order to obtain our results, it has been crucial the use of an
intrinsic method in order to find the characteristic speeds and
the corresponding eigenfields. This method, based on a General
Relativistic analogue of the well known Riemann problem of
Computational Fluid Dynamics (CFD), is completely independent of
the space coordinate system used at any given $t=constant$
hypersurface. It is presented in Section 3 and then applied in
Section 4 to a new simple one-parameter family of hyperbolic
systems in order to illustrate our arguments.

\section{A first order evolution system}

The evolution system (\ref{evolve_metric},\ref{evolve_curv}) is of
first order in time but second order in space. To obtain a system
which is also of first order in space, we will follow the standard
procedure by considering the first space derivatives of the metric
coefficients $\gamma_{ij}$ as new independent variables:
\begin{equation}\label{Ds}
    D_{kij} = 1/2\; \partial_k \gamma_{ij} \;\;.
\end{equation}
The evolution equation of these new variables $D_{kij}$ can then
be taken to be:
\begin{equation}\label{evolve_Ds}
    \partial_t D_{kij} + \partial_k(\alpha\; K_{ij}) = 0
\end{equation}
where we are using for simplicity normal coordinates, so that
$\beta^i=0$. The relationship (\ref{Ds}) can then be understood as
a first integral of the extended system (\ref{evolve_metric},
\ref{evolve_curv}, \ref{evolve_Ds}).

In order to complete this system, however, we need to provide
evolution equations for the lapse function $\alpha$ and its first
derivatives. It can be done in a straightforward way as follows:
\begin{eqnarray}
    \partial_t \ln \alpha = -\alpha\;Q 
\label{evolve_lapse} \\
    A_i = \partial_i \ln \alpha 
\label{As} \\
    \partial_t A_i + \partial_i(\alpha\;Q) = 0
\label{evolve_As1}
\end{eqnarray}
where, again, the equation (\ref{As}) must be considered as a
first integral of (\ref{evolve_As1}). Notice that one could
instead prescribe a more general evolution equation for $A_i$:
\begin{equation}\label{evolve_As2}
    \partial_t A_i + \partial_i(\alpha\;Q)
     + \mu\;\alpha\;(\nabla_k\:{K^k}_{i}-\partial_i\:trK) = 0
\end{equation}
where $\mu$ is an arbitrary parameter. In that case, (\ref{As})
would be a first integral of (\ref{evolve_As2}) if and only if the
momentum constraint (\ref{momentum_constraint}) is satisfied.

The complete first order evolution system can then be expressed in
a balance law form, namely:
\begin{equation}\label{balance_law}
    \partial_t u + \partial_k F^k =  S  \;,
\end{equation}

where $u$ is the array of independent variables (the metric
coefficients and their first derivatives), and both the Fluxes
$F^k$ vector array and the Sources $S$ scalar array depend of u.

Note however that the actual form of the fluxes associated to
$K_{ij}$ in eq. (\ref{evolve_curv}), will depend on the expression
of the three-dimensional Ricci tensor $^{(3)}R_{ij}$ in terms of the
metric derivatives (see for instance Ref. \cite{Cornell01} for a
discussion of this point). We will take in this paper the original
expression, namely:
\begin{equation}\label{Ricci3D}
   {^{(3)}R}_{ij}= \partial_k{\Gamma^k}_{ij} -\partial_j{\Gamma^k}_{ki}
                 + {\Gamma^k}_{kr}\;{\Gamma^r}_{ij}
                 - {\Gamma^k}_{ri}\;{\Gamma^r}_{kj}
\end{equation}

This is in contrast with the actual choice in most of the
hyperbolic formalisms published up to now, where a
three-dimensional version of the decomposition (\ref{Ricci4D}) is
used instead of (\ref{Ricci3D}) in order to mimic the properties
of generalized wave equations.

\section{The general relativistic Riemann problem}

The balance law form (\ref{balance_law}) of the evolution system
is familiar from the Computational Fluid Dynamics (CFD) domain,
where we can consider (\ref{balance_law}) in the sense of
distributions, so that discontinuous solutions for $u$ are allowed
(weak solutions). The discontinuity surface $\Sigma$ can be given
by:
\begin{eqnarray}
   \phi(x^i,t)=constant \nonumber\\
   \partial_t \phi + v^i\;\partial_i \phi = 0 \nonumber\;\;,
\end{eqnarray}

where $v^i$ is the velocity of the $x^i=constant$ points of the
surface $\Sigma$. The consistency of these weak solutions with
(\ref{balance_law}) requires the exact cancellation of the Dirac
$\delta$ terms arising where the derivatives of discontinuous
functions are performed. One gets then the well known
Rankine-Hugoniot conditions:
\begin{equation}\label{RH}
     v[u] = n_k[F^k]
\end{equation}
where the square bracket stands for the jump of the quantities
across $\Sigma$, $n_k$ is the unit normal to $\Sigma$ and $v\equiv
v^i\;n_i$ is the propagation speed of the discontinuity front. The
eigenvalue problem (\ref{RH}) is usually known in CFD as the
``Riemann problem''.

In General Relativity, we are familiar with 'matched' metrics
having discontinuous first order derivatives and other less common
weak solutions, like colliding waves. A peculiar feature in
General Relativity is however that the non-linear terms are
limited to the Sources $S$ in (\ref{balance_law}), which can not
originate $\delta$ terms and do not appear then in (\ref{RH}).
This means that the General Relativistic Riemann problem is a
linear one and can be solved explicitly. This implies also that
the eigenvalues $v$ of (\ref{RH}) will coincide with the
characteristic speeds, contrary to what happens in CFD, where one
finds a much richer structure so that, in addition to plain
discontinuities, one can get also shocks and rarefaction waves
propagating at non-characteristic speeds. We will take advantage
of these simplifications in what follows.

Let us then write down the Riemann problem corresponding to our
evolution system (\ref{evolve_metric}, \ref{evolve_curv},
\ref{evolve_Ds}, \ref{evolve_As2}), taking from granted that the
metric coefficients are continuous so that the only jumps can be
on the first derivatives quantities, namely
\begin{eqnarray}
    v[K_{ij}] &=& \alpha\;n_k[{\lambda^k}_{ij}]
\label{RH2_K} \\
    v[D_{kij}] &=& \alpha\;n_k[K_{ij}]
\label{RH2_D} \\
    v[A_i] &=& \alpha\;n_k[Q\;{\delta^k}_i +
                \mu\;({K^k}_i-trK\;{\delta^k}_i)]
\label{RH2_A}
\end{eqnarray}
where we have written, allowing for (\ref{Ricci3D}),
\begin{equation}\label{Fluxes_K}
   {\lambda^k}_{ij}=-{\Gamma^k}_{ij}
     + 1/2 \: \delta^k_i(A_j + D_j)
     + 1/2 \: \delta^k_j(A_i + D_i)
\end{equation}
and $D_k=\gamma^{ij}\;D_{kij}$.

Let us remark that the Riemann problem (\ref{RH2_K}-\ref{RH2_A})
is stated for a generic orientation of the unit normal $n_k$, in a
way that is independent of the space coordinate system. In this
sense, it provides an intrinsic method to obtain the
characteristic speeds and the corresponding eigenfields in terms
of geometrical objects related with both the $t=constant$
hypersurface and the characteristic surface $\Sigma$.

\section{Getting an strongly hyperbolic system}

We will now solve the eigenvalue problem (\ref{RH2_K}-\ref{RH2_A})
in order to display the causal structure of the evolution system
(\ref{evolve_metric}, \ref{evolve_curv}, \ref{evolve_Ds},
\ref{evolve_As2}). We remember that the system will be strongly
hyperbolic if and only if all the eigenvalues $v$ are real and if
the set of eigenfields is complete, in the sense that it spans all
our variable space. It follows easily from (\ref{RH2_D}) that
\begin{equation}\label{RH_Dt}
     v[D_{\perp ij}] = 0
\end{equation}
where the symbol $\perp$ replacing one index means that we are
taking the corresponding components tangent to $\Sigma$
(orthogonal to $n_k$). The remaining degrees of freedom in
(\ref{RH2_D}) can then be written as
\begin{equation}\label{RH_Dn}
     v[{D^n}_{ij}] = \alpha [K_{ij}]
\end{equation}
where we have noted for short ${D^n}_{ij} \equiv n_k\;{D^k}_{ij}$.

Let us now analyze the ``mixed'' components of (\ref{RH2_K})
\begin{eqnarray}\label{RH_Knt}
     v[{K^n}_{\perp}] &=& \alpha[-{D_{\perp}}^{nn}+1/2\;A_{\perp}] \\
     v[A_{\perp}] &=& \alpha\;\mu[{K^n}_{\perp}]
\end{eqnarray}
so that, allowing for (\ref{RH_Dt}), one gets the following set of
eigenvectors
\begin{equation}\label{eigenvectors1}
   \sqrt{\mu/2}\;{K^n}_{\perp} \pm (1/2\;A_{\perp}-{D_{\perp}}^{nn})
\end{equation}
with characteristic speeds $v=\pm\alpha\;\sqrt{\mu/2}$,
respectively. This means that our evolution system will be
(strongly) hyperbolic only if $\mu>0$. This surprising
parameter-dependence of the characteristic speeds can be
understood if we recall that physical solutions must verify the
momentum constraint (\ref{momentum_constraint}). For a
discontinuous solution one gets
\begin{equation}\label{RH_momentum}
   [{K^n}_{i}]=n_i[trK]
\end{equation}
so that the first term of (\ref{eigenvectors1}) comes from the
transverse part of (\ref{RH_momentum}). This means that the
characteristic cones spanned by (\ref{eigenvectors1}) are just
mathematical artifacts in order to get an hyperbolic system: they
can not propagate physical information. We will take $\mu=2$ to
impose this arbitrary characteristic speed to coincide with
light speed.

Let us focus now on the purely transverse components in
(\ref{RH2_K}). We get
\begin{eqnarray}\label{RH_Ktt}
     v[K_{\perp\perp}] &=& \alpha[-{\Gamma^n}_{\perp\perp}]=
          \alpha[{D^n}_{\perp\perp}-2\;{D_{(\perp\perp)}}^n] \\
     v[{D^n}_{\perp\perp}] &=& \alpha[K_{\perp\perp}]
\end{eqnarray}
so that, allowing again for (\ref{RH_Dt}), one gets the following eigenfields
\begin{equation}\label{eigenvectors2}
     K_{\perp\perp} \pm ({D^n}_{\perp\perp}-2\;{D_{(\perp\perp)}}^n)
\end{equation}
with characteristic speed $v=\pm \alpha$, independently of the
choice of $\mu$. Notice that the longitudinal component of
(\ref{RH_momentum}) can be written as
\begin{equation}\label{RH_momentum_n}
   (\gamma^{ij}-n^i\;n^j)[K_{ij}]=0
\end{equation}
so that, again, the trace component in (\ref{eigenvectors2}) can
not describe the propagation of any physical information. See for
instance Ref. \cite{Cornell01} to verify that the corresponding
characteristic speed can be also modified at will by using the
energy constraint in a suitable way. We can conclude that only the
traceless part of the transverse components (\ref{eigenvectors2})
can actually propagate physical information.These two degrees of
freedom are then good candidates to describe intrinsic features,
like gravitational waves, as we will discuss further in the next
section.

The remaining degrees of freedom in the original system
(\ref{RH2_K}-\ref{RH2_A}) are given by:
\begin{eqnarray}\label{gauge_cone}
     v[trK] &=& \alpha[A^n + 2 \;(D^n-{D_k}^{kn})] \\
     v[A^n] &=& \alpha[2\;(K^{nn}-trK) + Q] \\
     v[\gamma^{ij}\;{D^n}_{ij}] &=& \alpha[trK]
\end{eqnarray}
where now we must specify the gauge-related function $Q$ in terms
of other quantities. We have postponed the gauge specification up
to that point to emphasize that our previous analysis is gauge
independent. Now we will take the simple prescription
\begin{equation}\label{Q}
    Q= f\;trK
\end{equation}
where $f$ can be again an arbitrary parameter. We get then,
allowing once more for (\ref{RH_Dt}):
\begin{equation}\label{eigenvectors3}
     v[A^n + 2\;(D^n-{D_k}^{kn}) - f\;D^n] = 0
\end{equation}
and also that the following gauge-dependent combinations
\begin{equation}\label{eigenvectors4}
     \sqrt{f}\;trK \pm (A^n + 2\;D^n - 2\;{D_k}^{kn})
\end{equation}
are eigenfields with characteristic speeds
$v=\pm\alpha\;\sqrt{f}$, so that the system can be strongly hyperbolic
only if $f>0$. We can take for instance $f=1$ if we insist in having
light speed as a characteristic speed also here.

The analysis is complete now, because for $\mu=2,f>0$, the
eigenfields (\ref{RH_Dt}), (\ref{eigenvectors1}),
(\ref{eigenvectors2}), (\ref{eigenvectors3}) and
(\ref{eigenvectors4}) span the whole space of variables. It
follows that the resulting evolution system is strongly
hyperbolic, no matter of the particular orientation $n_k$ of the
characteristic surface $\Sigma$.

\section{Geometrical and physical interpretation}

Let us now decompose the line element of the three-dimensional
$t=constant$ hypersurfaces in the following way (2+1
decomposition):
\begin{eqnarray}\label{metric3D}
    \gamma_{ij}\;&dx^i&\;dx^j = -N^2\;dz^2 \nonumber \\
    &+& \sigma_{ab}\;(dx^a+\lambda^a\;dz)\;(dx^b+\lambda^b\;dz) \;\;\;a,b=1,2
\end{eqnarray}
where the coordinates $x^a$ span the characteristic surface
$\Sigma$, $\sigma_{ab}=\gamma_{ab}$ is the induced metric on
$\Sigma$ and $z$ is a transverse coordinate so that the unit
normal to $\Sigma$ can be expressed as
\begin{equation}\label{normal}
    n_i = N\;{\delta_i}^z \;\;.
\end{equation}
The eigenfields (\ref{eigenvectors2}) can then be written as
\begin{equation}\label{eigenvectors2_}
       K_{ab} \pm N\;({D^z}_{ab}-{D_{ab}}^z-{D_{ba}}^z)
\end{equation}
and a straightforward calculation shows that the extrinsic
curvature of the two-dimensional surface $\Sigma$, namely
\begin{equation}\label{curv2D}
    \kappa_{ab}\equiv
      1/(2N)\;[\partial_z\sigma_{ab}-{\cal L}_{\lambda}(\sigma_{ab})]
\end{equation}
is given by
\begin{equation}\label{curv2D_}
    \kappa_{ab}=N\;({D^z}_{ab}-{D_{ab}}^z-{D_{ba}}^z)
\end{equation}
so that the eigenfields (\ref{eigenvectors2}) can be expressed as
\begin{equation}\label{eigenvectors2_f}
    K_{ab} \pm \kappa_{ab}
\end{equation}
which shows that they are two-dimensional tensors associated to
the characteristic surface $\Sigma$ in an intrinsic,
coordinate-independent way.

Looking for a physical interpretations, let us note that the
traceless part of (\ref{eigenvectors2_f}) contains the shear
degrees of freedom as measured by a two-dimensional array of
observers sitting on the surface $\Sigma$. This is precisely the
kind of effect one expects from gravitational waves with wavefront
$\Sigma$. Our results can then be interpreted as an extension of
the standard description of the effect of a gravitational wave on
an array of freely falling (geodesic, $\alpha=constant$)
observers. This is an extension in the sense that our gauge
conditions (\ref{Q}) apply to more general kinds of observers
(like static ones, where both $Q$ and $trK$ vanish separately so
that (\ref{Q}) holds for any value of $f$). This can be relevant
for modelling Earth-based gravitational wave detectors, which are
certainly not in free fall.

Notice, however, that the eigenfields expression
(\ref{eigenvectors2}) is sensitive to the ordering ambiguity of
space derivatives when passing from (\ref{evolve_curv}) to a first
order system (see Ref. \cite{Cornell01} for details). The tensor
character of the eigenfields is completely lost if one starts with
a three-dimensional version of (\ref{Ricci4D}) instead of
(\ref{Ricci3D}). The intrinsic information resides of course in
the system (\ref{balance_law}), where the divergence of the fluxes
is taken, but the local covariance of the fluxes themselves is
lost if we make a choice different of (\ref{Ricci3D}). We claim
that our results provide a criterion for solving the ordering
ambiguity: the eigenfields describing the gravitational radiation
degrees must have an intrinsic geometrical meaning independent of
the space coordinate system. This opens the way to a
coordinate-independent local description of gravitational wave
detectors at Earth-based laboratories.

\section{Conclusions and outlook}

We have started this paper by proposing a new simple method, based
on the General Relativistic Riemann problem, in order to analyze
the causal structure of any first order flux-conservative
evolution system in an intrinsic way, independent of the space
coordinates. To compare with previous results \cite{Bona95}, we
can look at the characteristic speeds and see how we get now light
speed as $v=\pm \alpha$ (length per unit coordinate time), whereas
in coordinate-dependent analysis, the speed along the z axis is
given by  $v=\pm \alpha\;\sqrt{\gamma^{zz}}$ (coordinate
displacement per unit coordinate time).

The use of this intrinsic method has allowed ourselves to explain
the appearance of arbitrary parameter-dependent characteristic
speeds \cite{Cornell01} as a direct consequence of the mixing of
evolution and constraint equations. As we emphasized in Section
IV, only two degrees of freedom are intrinsically related with
causal propagation of physical information. Hyperbolic formalisms,
which can only consist of characteristic cones and lines, apply
instead to an extended, unconstrained, space of solutions. The
subspace of physical solutions is recovered by restricting the
initial conditions to  those that satisfy the constraints. The
extra degrees of freedom are then to be regarded as mathematical
artifacts devised to ensure hyperbolicity and their corresponding
(arbitrary) characteristic speeds are irrelevant from the physical
point of view.

Regarding the two remaining degrees of freedom, that can
consistently be interpreted as describing gravitational radiation,
we have provided an intrinsic geometrical interpretation of the
corresponding eigenfields in terms of the shear of the congruence
of eulerian (laboratory) observers combined with the shear of the
front-wave surfaces $\Sigma$. The requirement that the eigenfields
admit this geometrical interpretation allows one to solve the
ordering ambiguity for space derivatives when passing from the
second order form of Einstein evolution equations to a first order
system.

The fulfillment of this requirement must be regarded as an
important benchmark to check any hyperbolic formalism. This is
actually a strong requirement: none of the hyperbolic formalisms
proposed previously by the Palma group \cite{Bona92,Bona95,Bona97}
verifies it. It is not difficult, however, to modify the existing
formalisms to comply with it \cite{New}.

This ``explicit gravitational radiation'' requirement is mandatory
if we want to provide a local description of gravitational
radiation in order to model for instance the process of detection
of these waves by a given array of observers. In this sense, our
results generalize the well-known textbook results for freely
falling observers, based in the geodesic deviation equations. In
particular,  our results can be useful to model Earth-based
laboratories, where the Earth gravitational field can be
approximated by the well known Schwarzchild solution  and the
other effects, including the gravitational
wave itself, can be included using perturbation theory.\\

{\em Acknowledgements: We thank Manuel Tiglio for helpful and
stimulating discussions during his visit to Palma de Mallorca.
This work has been supported by the EU
Programme 'Improving the Human Research Potential and the
Socio-Economic Knowledge Base' (Research Training Network Contract
(HPRN-CT-2000-00137) and by a grant from the Conselleria
d'Innovacio i Energia of the Govern de les Illes Balears}

\bibliographystyle{prsty}

\end{document}